\documentclass[12pt,epsf]{article}
\newcommand{\NN}{{\cal N}}

\newcommand{\be}{\begin{equation}}
\newcommand{\ee}{\end{equation}}
\newcommand{\ben}{\begin{eqnarray}\displaystyle}
\newcommand{\een}{\end{eqnarray}}
\newcommand{\refb}[1]{(\ref{#1})}
\newcommand{\p}{\partial}

\newcommand{\al}{\alpha}

\newcommand{\s}{\sigma}
\newcommand{\la}{\lambda}

\newcommand{\Tr}{\hbox{Tr}}

\newcommand{\sectiono}[1]{\section{#1}\setcounter{equation}{0}}

\begin{document}

{}~ \hfill\vbox{\hbox{hep-th/0209047}
\hbox{UUITP-06/02}}\break

\vskip 1.cm

\centerline{\large \bf Circular Semiclassical 
String Solutions on $AdS_5\times S_5$}
\vspace*{1.5ex}

\vspace*{4.0ex}

\centerline{\large \rm Joseph A. Minahan\footnote{E-mail:
joseph.minahan@teorfys.uu.se}}
\vspace*{2.5ex}
\centerline{\large \it Department of Theoretical Physics}
\centerline{\large \it Box 803, SE-751 08 Uppsala, Sweden}
\vspace*{3.0ex}

\vspace*{4.5ex}
\medskip
\bigskip\bigskip
\centerline {\bf Abstract}
We discuss two semiclassical string solutions on $AdS_5\times S_5$.  In
the first case, we consider a multiwrapped circular string pulsating
in the radial direction of $AdS_5$, but fixed to a point on the $S_5$.
We compute the energy of this motion as a function of a large quantum
number $n$.  We identify the string level with $mn$, where $m$ is the 
number of string wrappings. 
  Using the AdS/CFT correspondence, we argue that the bare dimension
of the corresponding gauge invariant operator is $2n$ and that its
anomalous dimension scales as $\la^{1/4}\sqrt{mn}$, for large $n$.
Next we consider a multiwrapped
circular string pulsating about two opposite poles
of the $S_5$.  We compute the energy of this motion as a function of
a large quantum number, $n$  where again the string level is
given as $mn$.  We find that the dimension of the corresponding operator is 
$2n(1+f(m^2\la/(2n)^2))$, where $f(x)$ is  computible as a  series about
$x=0$ and where it is analytic.  We also compare this result to the BMN result
for large $J$ operators.

\bigskip

\vfill \eject
\baselineskip=17pt

\sectiono{Introduction}

The AdS/CFT correspondence identifies a large $N$ $\NN=4$ super Yang-Mills
theory with string theory propagating on an $AdS_5\times S_5$ background
\cite{9711200,9802150,9802109}.  
In order to verify the correspondence, one would hope
to solve string theory in this background, or at least be able to find
the spectrum of the free string states.  Major progress along this
direction was made in \cite{0112044},\cite{0202109} 
where it was shown that for a certain
plane wave limit of $AdS_5\times S_5$ \cite{0202111,0201081,0110242}, 
the spectrum of the string states
is exactly solvable.  Given this spectrum for the
string states,   Berenstein, Maldacena and Nastase (BMN) showed how to 
identify these particular
string states with gauge invariant operators with
large $R$-charge $J$
\cite{0202021} .  The energies
of the string states should be directly related to the dimensions of the 
operators, and the authors in \cite{0202021} showed that the identification was
consistent, at least to one loop level.   In \cite{0205066} it was shown
that the identification was consistent with a two loop calculation and,
modulo an assumption,
all higher loops as well.  In \cite{0206079}, using a very elegant
argument, it was shown that the BMN result was consistent with
all planar contributions.   It has also been recently shown
how to match up string loop
corrections with nonplanar diagrams for   large $J$ operators with
two scalar impurities
\cite{0205033},\cite{0205048},\cite{0205089}, as well as operators
with derivative impurities \cite{0208041},\cite{0208148} \footnote{
 However there are
also subtleties in matching higher point functions 
to the  full string field theory results of
\cite{0204146}-\cite{0206221}
because of mixings between single and double trace operators 
\cite{0205321}-\cite{0209002}.}. 

However, the plane wave  limit of BMN 
just describes one particular corner of the
complete set of gauge invariant operators in the field theory, so it would
be nice to be able to make some statements for string motion in 
$AdS_5\times S_5$ that does not correspond to this large $J$  limit.

The authors in \cite{0204051} described one way of doing this.  
Their idea was to consider the Nambu-Goto action for string motion
in $AdS_5\times S_5$, with the string motion limited to highly symmetrical
motion.  The main example that they discussed was a spinning string
stretched in the $AdS_5$ space.  They considered the classical motion
for this string, which by its symmetry, is solvable via quadratures.
If the stretching of the string is very small, then the string does not
really see the curvature of the space, so the results should be almost
identical to motion in flat space.  For such motion, they find a relation
between the energy of the string and its spin which is
\be\label{spdim}
E^2=2(\la)^{1/2}S,
\ee
where $\la$ is the 't Hooft coupling.  Since the dimension of the
 operator is equal to the energy of the dual string state $E$,
 one finds the result
\be
\Delta\ =\ (2\sqrt{\la}S)^{1/2}.
\ee
The corresponding operator is
\be
\Tr(\phi^I(\nabla^+)^{S}\phi^I),
\ee
where $\phi^I$ are the six adjoint fields and $\nabla^+$ is the covariant 
derivative
circularly polarized in the plane that the string is spinning in.  This
result is consistent with \cite{9802109} where it was argued that the
dimension for an operator corresponding to an excited string state is
\be
\Delta\ =\ 2(\sqrt{\la}n)^{1/2},
\ee
where $n$ is the level of the string state.  For the spinning string,
the level is $S/2$.

However, for large enough $S$, there must be some change in the behavior
of $\Delta$.  This is because the bare dimension is increasing linearly
in $S$, while \refb{spdim} is only increasing as $\sqrt{S}$.  If this
formula were to hold for all $S$, then at some point the bare dimension
of the operator would be greater than the full dimension, violating
unitarity.  

The resolution of this problem is that for large enough $S$, the string
starts seeing the curvature of the space, which will begin to modify 
$\Delta$.  In particular, for very large $S$, the authors in \cite{0204051}
found that 
\be
\Delta\ = \ S\ + \frac{\sqrt{\la}}{\pi} \log S.
\ee
Remarkably, the $\log S$ behavior appears in perturbative Yang-Mills
computations \cite{Gross}, albeit with a factor of $\la$.
Further study of this string motion was carried out in 
\cite{0204226}-\cite{0206103}

The spinning string analysis in \cite{0204051} was essentially semiclassical,
since it was assumed that the spin is quantized.  It is then clear
how to relate the level of the string state to the spin.
The purpose of this note is to consider other string motion where
a semiclassical analysis can be done, giving us a means to compute
an energy and hence a dimension of an operator, and then relate this
dimension to a string level.

We will consider two types of symmetric motion.  The first type
corresponds to a
multiwrapped circular pulsating 
string expanding and contracting in the $AdS_5$ subspace.  One
can then do a WKB analysis and relate a quantum number, namely the excitation
level of the string, to a dimension of an operator.  This quantum number
should be twice the level of the corresponding string state.  In this
case, we will find that the anomolous dimension of the corresponding gauge
theory operator behaves as
\be
\Delta\ -\ \Delta_b\ \sim\ (m\sqrt{\la}n)^{1/2}
\ee
for large $n$,
where $\Delta_b=2n$ is the bare dimension of the operator, $m$ is the
number of times the string wraps  over itself and $mn$ corresponds
to the string level. 

We then consider a circular pulsating string expanding and contracting on the $S_5$.
For very large quantum numbers, the string easily oscillates by expanding about
the north pole and shrinking about the south pole and then going back again.
In this case, we find that
\be
\Delta\ -\ \Delta_b\ \sim \ \frac{m^2\la}{2n}.
\ee
Hence, the anomalous dimension is analytic about $m^2\la/(2n)^2=0$, which seems to
indicate that this result can be directly verified in perturbation theory.
We also find that $m^2\la/(2n)^2$ acts as an expansion parameter, which is
similar to the BMN case where $m^2\la/J^2$ is the expansion parameter.
Since $m$ is closely related to relative phases in the BMN case
\cite{0202021},
we conjecture that the same is true for the dual operators considered here.
Using this assumption we are led to conjectures for the form of the operators
in both cases considered here.

In section 2 we start by considering a circular pulsating string in
flat space.  In section 3 we consider the string pulsating in the
$AdS_5$ space.  Using a WKB analysis we compute the energy of the
string state as a function of a large quantum number.  From our results in
the previous section, we are able to relate this quantum number to the 
string level.    In section 4 we consider the string pulsating in
the $S_5$.  Using ordinary perturbation theory, we are able to compute
corrections to the energies for large quantum numbers.  We also argue
that the operator dual to the string state approaches a chiral primary
as $n$ becomes large.
In section 5 we give our conclusions.

\sectiono{The pulsating string in flat space}

As a warmup, let us consider quantization of a circular pulsating string
in flat space.  We will assume that all motion is in a two dimensional
plane and so the relevant metric is
\be
ds^2\ =\ -dt^2\ +\ dr^2\ +\ r^2\ d\theta^2.
\ee
The Nambu-Goto action is
\be\label{NGflat}
S\ = \ -\frac{1}{2\pi\al'}\int d\tau d\s
\sqrt{-\det(\p_\al X^\mu\p_\beta X_\mu)}.
\ee
This
can be reduced by invoking the reparameterization invariance to set
$t=\tau$ and $\theta=m\s$, where we allow for a string
that is multiwrapped.  Assuming that the radial coordinate has
no angular dependence, the action in \refb{NGflat} becomes
\be
S\ = \ -\frac{m}{\al'}\int dt\ r\ \sqrt{1-\dot r^2}.
\ee
The canonical momentum is
\be
\Pi\ =\ \frac{r\dot r}{m\al'\ \sqrt{1-\dot r^2}},
\ee
and so the Hamiltonian is
\be
H\ =\ \Pi\ \dot r\ -\ L\ =\ \sqrt{\Pi^2\ +\ m^2r^2/\al'^2}.
\ee
Hence the square of the Hamiltonian looks like an ordinary harmonic
oscillator, although constrained to the even states since $r$ is
a radial component.  Hence, for large energy states, the energy squared
satisfies\footnote{We are ignoring the zero point corrections.  They
are unimportant for the discussion contained here.} 
\be
(E_n)^2\ =\ 2(2nm)/\al',
\ee
where $n$ is an integer.   The product $nm$ is the level.

It is simple to see where these states come from in the more canonical
way of treating closed strings in flat space.  The solutions for the
classical equations of motion are
\be
X^\mu(\tau,\s)\ =\ p^\mu\tau\ +\ x^\mu\ +\ \sum_{n\ne0}\frac{\al^\mu_{-n}}{n}
\  e^{in(\tau-\s)}+\frac{\widetilde\al^\mu_{-n}}{n}\  e^{in(\tau+\s)},
\ee
with the constraints
\be
\dot X^\mu\dot X_\mu+{X'}^\mu{X'}_\mu=0\qquad\qquad \dot X^\mu{X'}_\mu=0.
\ee
Clearly, one solution is
\be\label{sol1}
X^1\ =\ A\cos m\tau\cos m\s\qquad\qquad X^2\ =\ A\cos m\tau\sin m\s,
\ee
which corresponds to 
\be
\widetilde\al^1_{-m}=\al^1_{-m},\qquad \al^2_{-m}=i\al^1_{-m},\qquad 
\widetilde\al^2_{-m}=-i\al^1_{-m},
\ee
along with similar relations for the complex conjugate variables.  All other
$\al_{-n}^\mu$ are zero.
In terms of the components $X^{\pm}=X^1\pm X^2$, we have
\be
\al^+_{-m}=\widetilde \al^-_{-m}\qquad\qquad\al^-_{-m}=\widetilde \al^+_{-m}=0.
\ee
Hence, for the quantized string,
the level $nm$ state corresponding to the circular string state
is
\be
(\al^+_{-m})^n(\widetilde\al^-_{-m})^n|0\rangle,
\ee
which clearly has $E^2=4nm/\al'$.

There is another circular state in this plane, which comes from interchanging
the $+$ and $-$ polarizations between the left and right movers.  In terms
of the solutions in \refb{sol1}, this corresponds to the transformation
$X^2\to -X^2$, which is a parity inversion.

We can compare this to the solution of the spinning string.  In this case
we have that
\be\label{sol2}
X^1\ =\ A\cos\tau\cos\s\qquad\qquad X^2\ =\ A\sin\tau\cos\s,
\ee
which corresponds to 
\be
\widetilde\al^1_{-1}=\al^1_{-1},\qquad \al^2_{-1}=i\al^1_{-1},\qquad 
\widetilde\al^2_{-1}=i\al^1_{-1},
\ee
and so
\be
\al^+_{-1}=\widetilde \al^+_{-1}\qquad\qquad\al^-_{-1}=\widetilde \al^-_{-1}=0.
\ee
The corresponding quantum state is
\be
(\al^+_{-1})^n(\widetilde\al^+_{-1})^n|0\rangle.
\ee
It was argued in \cite{0204051} that this string state is dual to the twist
2 operator
$\Tr(\phi^I(\nabla^+)^{2n}\phi^I)$.  

Hence, it is natural to expect that the pulsating
string corresponds to the trace of $n$ $\nabla^+$ operators and $n$ $\nabla^-$
operators sandwiched between two adjoint fields.  Unlike, the spinning string
where the covariant derivatives are automatically symmetric, we expect
that for the pulsating case, it will be necessary to symmetrize over
all orderings of $\nabla^+$ and $\nabla^-$.  We should also have
to include some $m$ dependent phases.  Also, the opposite parity string
state should correspond to interchanging $\nabla^+$ with $\nabla^-$.

\sectiono{The pulsating string in $AdS_5$}

In this section we consider the semiclassical quantization of a circular
string that expands and contracts in $AdS_5$.  The classical motion
for this was first considered in \cite{9410219,0204051}.  The Nambu-Goto action
is
\be\label{NGaction}
S=-\frac{1}{2\pi\al'}\ \int d\tau d\s\ \sqrt{-\det(G_{\mu\nu}\p_\al X^\mu\p_\beta X^\nu)},
\ee
where $G_{\mu\nu}$ is the $AdS_5$ metric
\be
ds^2\ =\ R^2(-\cosh^2\rho\ dt^2\ +\ d\rho^2\ +\sinh^2\rho\ d\Omega_3^2),
\ee
and where $R^4=\la{\al'}^2$ with $\la$ the 't Hooft coupling.
As usual, it is convenient to parameterize the 3-sphere metric as
\be
d\Omega_3^2= \cos^2\theta\ d\psi^2\ +\ d\theta^2\ +\ \sin^2\theta\ d\phi^2.
\ee
We will look for solutions with $\theta=\pi/2$ and the string wrapped $m$
times around
the $\phi$ direction.  Using the reparamerterization invariance of \refb{NGaction},
we identify $t=\tau$ and $\phi=m\s$.  Hence the action reduces to
\be\label{NGred}
S=-m\sqrt{\la}\ \int dt\ \sqrt{\cosh^2\rho-\dot\rho^2}\ \sinh\rho.
\ee

To proceed, it is convenient to choose a new variable $\xi$ whose relation
to $\rho$ is 
\be
\xi\ =\ \sin^{-1}(\tanh\rho).
\ee
Using the $\xi$ variable, the action becomes
\be\label{NGredn}
S=-m\sqrt{\la}\ \int dt\ \tan\xi\sec\xi\sqrt{1-\dot\xi^2}.
\ee
The canonical momentum is given by
\be\label{cmom}
\Pi\ =\ m\sqrt{\la}\tan\xi\sec\xi\frac{\dot\xi}{\sqrt{1-\dot\xi^2}},
\ee
and so the Hamiltonian is
\be\label{ham}
H\ =\ \Pi\ \dot\xi-L\ =\ \sqrt{\Pi^2+m^2\la\tan^2\xi\sec^2\xi}.
\ee
Hence, we see that $\frac{1}{2}H^2$ is an ordinary one
dimensional quantum mechanical system with potential
\be\label{pot}
V(\xi)=2m^2\la\tan^2\xi\sec^2\xi.
\ee

The potential in \refb{pot} is zero at $\xi=0$ and diverges at $\xi=\pi/2$. 
Hence for states with large energies, the potential is approximately that
of a square well, with $\xi$ constrained to be positive.  Alternatively,
we can consider the case where $\xi$ ranges from $-\pi/2$ to $\pi/2$, but
only wave-functions even in $\xi$ are considered.  That is the approach
that we will take here.  For large energy states, we then find that
the approximate wave functions are
\be
\psi_n(\xi)\ \approx\ \cos((2n+1)\xi),
\ee
and the approximate energies are \cite{9410219}
\be
E^2\approx (2n+1)^2.
\ee
Hence, we find that the dimensions of the operators approach
\be
\Delta\ \approx\ 2n+1,
\ee
which is the bare dimension of the operator.

However, we can do better than this and make a prediction for the leading
order correction for the energy of the string state, 
which would correspond to the anomalous
dimension of the operator.  We can do this by doing a Bohr-Sommerfeld
analysis for the quantization of the states.  Since we are only looking
for even wave functions, the quantized states satisfy
\be\label{BS}
(2n+1/2)\pi\ =\ \int_{-\xi_0}^{\xi_0}\sqrt{E^2-m^2\la\tan\xi\sec\xi}\ +\ {\rm O}(n^{-1}).
\ee
The values $\pm\xi_0$ are the turning points, which satisfy
\be
E=m\sqrt{\la}\tan\xi_0\sec\xi_0.
\ee
If we define $B=E/(m\sqrt{\la})$ and define a new variable $y=(B)^{-1/2}\tan\xi$,
then the integral in \refb{BS} can be written as
\be\label{BSint}
2m\sqrt{\la B}\left[\int_0^{y_0}\frac{dy}{B^{-1}+y^2}-
\int_0^{y_0}\frac{dy}{B^{-1}+y^2}\left(1-\sqrt{1-y^2/B-y^4}\right)\right].
\ee
The first integral inside the square brackets is
\be
\int_0^{y_0}\frac{dy}{B^{-1}+y^2}\ =\ \sqrt{B}\left(\frac{\pi}{2}
-\frac{1}{\sqrt{B}}\right).
\ee
The second integral is finite in the limit $B\to\infty$.  In this limit
the integral becomes
\be
\int_0^1\frac{dy}{y^2}(1-\sqrt{1-y^4})\ =\ -1\ +\ 
\frac{(2\pi)^{3/2}}{\Gamma\left(\frac{1}{4}\right)^2}.
\ee

Putting everything together, we see that for large values of $E$, the 
expression in \refb{BS} can be approximated as
\be
(2n+1/2)\pi\ \approx\ E\pi\ -\ \frac{4\pi(2\pi)^{1/2}}
{\Gamma\left(\frac{1}{4}\right)^2} m^{1/2}\la ^{1/4}\sqrt{E}.
\ee
Inverting this we find
\be\label{dim}
\Delta\ =\ E\ \approx\ 2n\ 
+\ \frac{8\pi^{1/2}}{\Gamma\left(\frac{1}{4}\right)^2}
\la ^{1/4}\sqrt{nm}.
\ee
Note that the numerical coefficients are similar to those that appeared
in the Wilson loop computations of \cite{9803001},\cite{9803002}.

Hence, from \refb{dim} we see that this operator with a large bare
dimension has an anomalous piece which is of the same form as a nonchiral
operator with a low bare dimension.  For such an operator, 
the full dimension was found to be
\be
\Delta\ \approx\ 2\la ^{1/4}\sqrt{nm},
\ee
for level $nm$.
For the case described in \refb{dim} the prefactor is
\be
\frac{8\pi^{1/2}}{\Gamma\left(\frac{1}{4}\right)^2}\ \approx\ 1.0787.
\ee

Let us rewrite $\Delta$ in \refb{dim} as
\be\label{dim2}
\Delta\  \approx\ 2n\left(
1\ +\ \frac{8\pi^{1/2}}{\Gamma\left(\frac{1}{4}\right)^2}
\left(\frac{\la m^2}{n^2}\right)^{1/4}\right).
\ee
Hence, we see that $\frac{\la m^2}{n^2}$ acts as an expansion parameter,
which is quite similar to the expansion parameter in BMN.  In their case,
the role of $n$ is played by $J$, the value of the $R$-charge.  The role
of $m$ appears the same; it corresponds to the level of each oscillator.
Note further that this expansion parameter is multiplied by a factor of
$n$, suggesting that this state corresponds to one with $n$ right and left
moving oscillators.
We will see an even closer similarity to the BMN results in the next section.

Let us conclude this section by postulating an operator that is dual
to this string state.  The bare dimension should be $2n$ and the total
spin is zero.  The motion is entirely in the $AdS_5$, suggesting that
the operator is built out of a symmetrized sum over the
product of covariant derivatives.
 The analysis of BMN also suggests that the level of each oscillator
introduces a phase $e^{2\pi i m/n}$ when a $\nabla^{+}$ operator goes
through a $\nabla^{-}$ operator.   Hence, one might guess that the operator
is
\be
\frac{n!}{\sqrt{(2n)!}}
\sum_{perms}\Tr\left(\phi^I (\nabla^+)^{n^+_1}(\nabla^-)^{n^-_1}
(\nabla^+)^{n^+_2}(\nabla^-)^{n^-_2}...(\nabla^+)^{n^+_k}(\nabla^-)^{n^-_k}
\phi^I\right) \exp(i\varphi(n^+_1,n^-_1,...n^+_k,n^-_k)),
\ee
where the sum is over all orderings of the covariant derivatives.  We also
have that $k\le n$ and that the $n^+_i$ and $n^-_i$ satisfy
\be\label{nsums}
\sum_{i=1}^k n^+_i=\sum_{i=1}^k n^-_i=n.
\ee
The phase is given by
\be\label{phase}
\varphi(n^+_1,n^-_1,...n^+_k,n^-_k)\ =\ -\frac{2\pi m}{n}\sum_{i\le j}^kn^+_in^-_j.
\ee

Notice that the operator is not invariant under the exchange of
$\nabla^+$ with $\nabla^-$ because of the phase terms.  This fits with
our expectation that changing the orientation of the string leads to a new
string state.  Note further that the transformation of $m\to n-m$ is
equivalent to changing the orientation.  This operator must have the
same dimension as the original operator.  This seems at first to contradict
the result in \refb{dim2}.  However there is no contradiction, since
in order for the approximation to work, we must assume that
$\frac{\la m^2}{n^2}$ is small.  Since $\la$ is assumed to be large, if
$\frac{\la m^2}{n^2}$ is small then $\frac{\la (n-m)^2}{n^2}$ is
large.

\sectiono{The pulsating string on $S_5$}

As our other  example\footnote{This section has been expanded from the
earlier version.  I thank Arkady Tseytlin, for pointing out
an error in the previous version of this section.}, 
let us consider a circular pulsating string expanding
and contracting on $S_5$.  
Let us write the metric on $S_5$ as
\be\label{S5met}
ds^2\ =\ R^2(\cos^2\theta\ d\Omega_3^2\ +\ d\theta^2\ +\ \sin^2\theta\ d\psi^2).
\ee
We will assume that the string is stretched along $\psi$ and is at a fixed
point on the $S_3$ and the spatial components of the $AdS_5$.  Hence, the
relevant metric for us is
\be 
ds^2\ =\ -R^2(-\cosh^2\rho\ dt^2\ +\ d\rho^2\ +\ d\theta^2\ 
+\ \sin^2\theta\ d\psi^2).
\ee
We have chosen to keep the radial piece of the $AdS_5$ in order to account
for quantum fluctuations in this space.
Identifying $t$ with $\tau$ and $\psi$ with $m\s$ to again allow for
multiwrapping, the Nambu-Goto action
reduces to
\be\label{NGsph}
S\ =\ m\sqrt{\la}\int dt\ \sin\theta\ \sqrt{\cosh^2\rho-\dot\rho^2-\dot\theta^2}.
\ee
Hence, the canonical momentum are
\begin{eqnarray}
\Pi_\rho\ &=&\ \frac{m\sqrt{\la}\ \sin\theta\ \dot\rho}{\sqrt{\cosh^2\rho-\dot\rho^2-\dot\theta^2}},\\
\Pi_\theta\ &=&\ \frac{m\sqrt{\la}\ \sin\theta\ \dot\theta}{\sqrt{\cosh^2\rho-\dot\rho^2-\dot\theta^2}}.
\end{eqnarray}
Solving for the derivatives in terms of the canonical momenta
and substituting into the Hamiltonian, we find 
\be
H\ =\ \cosh\rho\ \sqrt{\Pi_\rho^2+\Pi_\theta^2+m\la\sin^2\theta}.
\ee
If we were to fix $\rho=0$, then the squared Hamiltonian looks very much like a
one dimensional pendulum.

For small energy states, $\sin\theta<<1$, so the squared Hamiltonian reduces
to the usual harmonic oscillator in flat space.  We are interested in
the opposite limit, where $E^2>>m^2\la\sin^2\theta$.  
The potential term comes from the stretching of the string, so if we ignore
it, we should expect particle like behavior.  The wave functions are
for
a free particle on $AdS_5\times S_5$, and are assumed to have
$\rho$ and $\theta$ dependence only.  Hence, they satisfy
\be
E^2\psi(\rho,\theta)\ =\ -\frac{\cosh\rho}{\sinh^3\rho}\frac{d}{d\rho}
\cosh\rho\sinh^3\rho\frac{d}{d\rho}\psi(\rho,\theta)\ -\ 
\frac{\cosh^2\rho}{\sin\theta\cos^3\theta}\frac{d}{d\theta}
\sin\theta\cos^3\theta\frac{d}{d\theta}\psi(\rho,\theta),
\ee
where we have taken into account the directions transverse to $\rho$ and $\theta$ in
expressing the momentum operators.
Hence, these
wave functions are
\be\label{wf}
\psi_{2n}(\rho,\theta)\ =\ (\cosh\rho)^{-2n-4}P_{2n}(\cos\theta),
\ee
with energies
\be
E_{2n}=\Delta=2n+4.
\ee 
The $P_{2n}(\cos\theta)$ are spherical harmonics on $S_5$, which
are even about $\theta=\pi/2$, assuming that the wave functions are
constant on the $S_3$.

To find the anomalous dimension, we can do ordinary perturbation
theory.  The wavefunctions are highly peaked around $\rho=0$ for large $2n$,
so we simply fix $\rho$ and only worry about the $\theta$ dependence.
We may also approximate the spherical harmonics as
\be
 P_{2n}(\cos\theta)\ \approx\ \sqrt{\frac{4}{\pi}}\cos 2n\theta.
\ee
Hence, the first term in the perturbative expansion gives
\be\label{E2S5}
\delta {E}^2\ =\ \int_0^{\pi/2} d\theta  \psi^*_{2n}(0,\theta)m^2\la\sin^2\theta
\psi_{2n}(0,\theta)\ =\ \frac{m^2\la}{2}.
\ee
Including a few higher order terms gives
\be
\delta {E}^2\ =\  \frac{1}{2}m^2\la\ + \frac{1}{32}\frac{m^4\la^2}{4n^2}
\ +\ {\rm O}(m^8\la^4),
\ee
where the third order perturbation turns out to be zero. 
Thus we find that the dimension of the corresponding operator is
\be\label{dimS5}
\Delta\ -4\ =\ 2n\left(1\ +\ \frac{1}{4}\left(\frac{m^2\la}{(2n)^2}\right)
-\ \frac{1}{64}\left(\frac{m^2\la}{(2n)^2}\right)^2\ 
+\ {\rm{O}}\left(\frac{m^2\la}{(2n)^2}\right)^3\right)
\ee

It is clear from \refb{dimS5} that again there is a natural expansion parameter
$\frac{m^2\la}{(2n)^2}$ which is similar to the  large $J$  
expansion parameter $\frac{m^2\la}{J^2}$.  In both cases, the term in
the denominator is the square of the bare dimension of the operator.
As in the BMN case, we see that even if $\la$ is large, the
expansion parameter can be chosen small by choosing $n$ large enough.
What is further significant about \refb{dimS5}, and unlike the case
in the previous section, is that the  anomalous piece is
analytic about $\frac{\la}{(2n)^2}=0$.  This suggests that there
exists a direct verification of this result in perturbation theory.

There is one other interesting similarity between  BMN and
this situation.  For BMN, the string state with $q$ $\al_{-m}$
and $q$ $\widetilde\al_{-m}$ oscillators has  $\Delta$ equal to
\be\label{largeD}
\Delta\ = \ J\ +\ 2q\ \sqrt{1+\frac{m^2\la}{J^2}}\ =\ J\ +
\ 2q\ \frac{1}{2}\ \frac{m^2\la}{J^2}
\ +\ {\rm O}\left(\left(\frac{\la}{J^2}\right)^2\right).
\ee
In \refb{largeD} $2q$ is the number of oscillators, $qm$ is the level and $J$ 
is the bare dimension, assuming
$q<<J$.  
In \refb{dimS5},
$2n$ is the number of oscillators,  $nm$ is the level and $2n$ is
the bare dimension.  So in this sense the linear
piece in
the two expressions matches, except that our result indicates
that there is a factor of 2 suppression because of the large number of
oscillators.

The corresponding operator for this string state would be expected to have
bare dimension $2n+4$, or $2n$, depending on whether or not $F^2$  is
inserted into the trace.  The chiral primary (corresponding to $m=0$)
should contain $2n$ scalar
fields in a symmetric representation of $SO(6)$.  The particular primary should
also be invariant under an $SO(4)\times U(1)$ subgroup of $SO(6)$.  So for example,
the primary without $F^2$ for $n=1$ is
\be
\Tr\left(\sum_{I=1}^4\Phi_I^2-\ 2Z\bar Z\right).
\ee

If $m\ne0$, then the corresponding operators are no longer chiral, but they should
be almost chiral in a manner similar to BMN.  We expect that this works by introducing
phases in the symmetrization over the scalar fields.  Since the string has no
excitations on the $S_3$ subspace of the $S_5$, we expect that the $SO(4)$ subgroup
is unbroken.  Hence, no phases should be introduced when exchanging the $\Phi_I$
amongst themselves.  But phases should be inserted when exchanging $Z$ with
$\bar Z$ or $Z$ with $\Phi_I$.  
This will also insure that the operator is not invariant under
$Z\to \bar Z$.

\sectiono{Conclusions}

In this paper, we discussed the semiclassical treatment for two examples
of highly symmetric string motion.  We learned that the curvature of
the background  appears to assure us of positive anomolous dimensions
when identifying energies of string states with dimensions of gauge
invariant operators.

In the case of motion on $S_5$ we found that the anomalous dimension is
analytic about $\la/(2n)^2=0$, which would seem to indicate that the results
can be directly compared with perturbation theory.  The fall off
is not as rapid as for BMN operators.  This is reasonable, since
we expect
the number of excited oscillators to be roughly the same as the
bare dimension.  Whether or not one can find such behavior in
a field theory computation is presently being investigated  \cite{jmkz}.

It would also be interesting to generalize these semiclassical
solutions 
to less symmetric motion.  For example, one can consider
doing small perturbations about the motion as was recently discussed
in \cite{0204226},\cite{0206103}.  This is presently being
studied.

\bigskip
\noindent {\bf Acknowledgments}:
I would like to thank I. Klebanov, A. Tseytlin and especially
 K. Zarembo for helpful
conversations and K. Zarembo for comments on the manuscript.
I would also like to thank the Aspen Center for Physics
and the CTP at MIT
 for hospitality during the course of 
this work.   This research was
supported in part by Vetenskapr\aa det.

\end{document}